\documentclass[12pt, a4paper]{article}
\usepackage{graphicx}
\usepackage{epsfig}
\usepackage{jcappub}
\usepackage{epstopdf}

\def \lleq {\lower0.9ex\hbox{ $\buildrel < \over \sim$} ~}
\def \ggeq {\lower0.9ex\hbox{ $\buildrel > \over \sim$} ~}

\def \beq  {\begin{equation}}
\def \eeq  {\end{equation}}
\def \ber  {\begin{eqnarray}}
\def \eer  {\end{eqnarray}}

\newcommand{\be}{\begin{equation}}
\newcommand{\ee}{\end{equation}}
\newcommand{\ba}{\begin{eqnarray}}
\newcommand{\ea}{\end{eqnarray}}
\newcommand{\bea}{\begin{eqnarray*}}
\newcommand{\eea}{\end{eqnarray*}}

\title{Constraints on cosmological parameters in power-law cosmology}
\author{Sarita Rani$^1$, A. Altaibayeva$^2$,  M. Shahalam$^3$, J. K. Singh$^1$, R. Myrzakulov$^2$}
\affiliation{$^1$Netaji Subhas Institute of Technology, University of Delhi, New Delhi, India}
\affiliation{$^2$Department of General and Theoretical Physics, Eurasian
National University,\\ Astana, Kazakhstan}
\affiliation{$^3$Center For Theoretical Physics, Jamia Millia Islamia, New Delhi, India}

\emailAdd {sarita\_{maths@yahoo.co.in}, aziza.bibol@mail.ru,  mdshahalam@ctp-jamia.res.in}
\emailAdd { jainendrrakumar@rediffmail.com, rmyrzakulov@gmail.com}

\abstract{ In this paper, we examine  observational constraints on the power law cosmology; essentially dependent on two parameters $H_0$ (Hubble constant) and $q$ (deceleration parameter). We investigate the constraints on these parameters using the latest 28 points of H(z) data and 580 points of Union2.1 compilation data and, compare the results with the results of $\Lambda$CDM. We also forecast constraints using a simulated  data set for the future JDEM, supernovae survey. Our studies give better insight into power law cosmology than the earlier done analysis by Kumar [arXiv:1109.6924] indicating it tuning well with Union2.1 compilation data but not with H(z) data. However, the constraints obtained on $<H_0>$ and $<q>$ i.e. $H_0$ average and $q$ average using the simulated data set for the future JDEM, supernovae survey are found to be  inconsistent with the  values obtained from the H(z) and Union2.1 compilation data. We also perform the statefinder
  analysis and find that the power-law cosmological models approach the standard $\Lambda$CDM model as $q\rightarrow -1$. Finally, we observe that although the power law cosmology explains several prominent features of evolution of the Universe, it fails in details.}

\date{\today}


\begin{document}

\maketitle

%\newpage
\section{Introduction}
\label{sect1}The Standard Cosmological Model (SM) of Universe {\it
a la} $\Lambda$CDM complemented by the inflationary phase is
remarkably a successful theory, although, the  cosmological constant
problem still remains to be one of the major unsolved
problems \cite{sami} of our times. It is therefore reasonable  to
examine the alternative cosmological models to explain the observed
Universe. Power-law cosmology is one of the interesting alternatives
to deal with some usual problems (age, flatness and horizon problems etc.) associated with the standard model. In
such a model, the cosmological
 evolution is explained by the geometrical scale factor $a(t) \propto t^\beta$ with $\beta$
 as a positive constant. The power law evolution with $\beta\geq1$ has been discussed
  at length in a series of articles in distinct
   contexts \cite{lohiya,batra1,batra2,geh1,geh2,dev1,dev2,sethi2,zhu}; phantom
power-law cosmology is discussed in reference \cite{kae}. The motivation
for such a scenario comes from a number of considerations. For
example, power-law cosmology does not face the horizon problem \cite{sethi2}, as well as the flatness problem. Another
remarkable feature of these models is that they easily accommodate
high redshift objects and hence reduce the age problem. These
models also deal with the fine tuning problem, in an attempt to
dynamically solve the cosmological constant problem
\cite{mann,allen,dol,ford,wein}.

A power law evolution
of the cosmological scale factor with $\beta \approx 1$ is  an
excellent fit to a host of cosmological observations. Any model
supporting such a coasting presents itself as a falsifiable model as
far as classical cosmological tests are concerned as it exhibits
distinguishable and verifiable features. Classical cosmological
tests also support such kind of evolution, such as the galaxy number
counts as a function of redshift  and the data on angular diameter
distance as a function of redshift \cite{kolb}. However,  these
tests are not considered as reliable tests  of a viable model since
these are marred  by evolutionary effects (e.g. mergers). Now,
SNe Ia (reliable standard candles), and hubble test have
 become more reliable to that of a precision measurement.

Cosmological parameters prove to be the backbone of any of the cosmological models, therefore it becomes important to obtain a concise range or more specifically, the estimated values of such parameters using available observational data, so that the said model can explain the present evolution of Universe more precisely. In this series of cosmological parameters we observe that Hubble constant ($H_{0}$) and
deceleration parameter ($q$) are very  important in describing the current nature of the Universe. $H_0$ explains the current expansion rate of the Universe whereas $q$ describes the nature of the expansion rate. In last few years, various attempts have been done to evaluate the value of  $H_{0}$. Freedman et al. \cite{free} evaluated a value of $H_{0}=72 \pm 8$ km/s/Mpc. Suyu et al.\cite{suy} evaluated  $H_{0}$ as $69.7_{-5.0}^{+4.9}$ km/s/Mpc. WMAP7  evaluated the value of $H_{0}=71.0\pm 2.5$ km/s/Mpc (with WMAP alone), and $H_{0} = 70.4_{-1.4}^{+1.3}$ km/s/Mpc (with Gaussian priors ) \cite{jar}. Numerous other evaluates of $H_{0}$ are  $73.8 \pm 2.4$ km/s/Mpc \cite{rie},  $67.0\pm 3.2$ km/s/Mpc \cite{beu}. Most recent PLANCK evaluate of the Hubble constant gives a value of  $H_{0}$ = $67.3\pm 1.2$ km/s/Mpc \cite{Ade:2013zuv}. Along with the above mentioned evaluates of $H_0$, several other authors, \cite{dev1}, \cite{sethi2,zhu,bgum,sur,gum}  obtained the constraints on  cosmological parameters including $H_0$, $q$ and $\beta$ for open, closed and  flat power law cosmology. Numerical results for flat power-law cosmology have been described in Table \ref{tabparm}.

In a recent paper, Kumar \cite{sur} has investigated observational constraints on the power-law cosmological parameters using H(z) and SN Ia data and discussed various features of power-law cosmology. In the present work, we are investigating the scenario similar to an analysis done in reference \cite{sur} for flat power law cosmology. We compare the results of the model under consideration with the results obtained from $\Lambda$CDM and with that of Kumar \cite{sur}. We use the most  recent observational datasets such as 28 points of H(z) data \cite{Farooq:2013hq} and  Union2.1 compilation (SN) data \cite{Suzuki:2011hu} (taking into account the full covariance matrix). Here, we also forecast constraints using a  simulated data set for the   future JDEM, supernovae survey \cite{hol,ald} and also employ Statefinder analysis of the results obtained.
%%%
\section{Power Law Cosmology}
For a flat FLRW metric, the line element is
\beq\label{p1}
 ds^2=c^2 dt^2-a^2(t)\left[dr^2+r^2(d\theta^2+sin^2\theta d\phi^2)\right],
\eeq
where $a(t)$ is the scale factor and $t$ is the cosmic proper time. In this paper, we discuss general power law cosmology,
\beq\label{p2}
a(t)=a_0\left(\frac{t}{t_0}\right)^{\beta}.
\eeq
Here, $t_0$ and $\beta$ represent the present age of Universe and dimensionless positive parameter respectively. Here and subsequently, the subscript $0$ defines the present-day value of the parameters considered.\\
$$H=\frac{\dot{a}}{a}=\frac{\beta}{t},$$ and $$H_0=\frac{\beta}{t_0}.$$\\
The relation between the red shift and the scale factor is given by\\
\beq\label{p3}
\frac{a(t)}{a_0}=\frac{1}{1+z}.
\eeq
The age of Universe at redshift $z$ is given as
\beq\label{p5}
t(z)=\frac{\beta}{H(z)},
\eeq
where,
\beq\label{p6}
H(z)=H_0 (1+z)^\frac{1}{\beta}.
\eeq
The acceleration of Universe can be measured through a dimensionless cosmological function called as the deceleration parameter $q$. In this scenario $q$ is
\beq\label{p7}
q=-\frac{\ddot{a}}{aH^2}=\frac{1}{\beta}-1,
\eeq
where, $q<0$ explains an accelerating Universe, whereas $q\geq0$ describes a Universe which is either decelerating or expanding at the 'coasting' rate.
Equation (\ref{p6}) in terms of $q$ can be written as
\beq\label{p8}
H(z)=H_0 (1+z)^{(1+q)}.
\eeq
Equation (\ref{p8}) implies that the parameters $H_0$ and $q$ explain history of the Universe in power law cosmology.
In this paper, we study the well behaved power-law cosmological model, focusing on the parameters $q$ and $H_0$, also we find the observational constraints on both of the above parameters to the latest $28$ data points of H(z) \cite{Farooq:2013hq} and Union2.1 compilation data \cite{Suzuki:2011hu} of 580 points. We also use the simulated data for upcoming Supernova (SN) surveys like JDEM to constrain the above said parameters \cite{hol,ald}.
%%%%
\begin{figure*} \centering
\begin{center}
$\begin{array}{c@{\hspace{0.4in}}c}
\multicolumn{1}{l}{\mbox{}} &
        \multicolumn{1}{l}{\mbox{}} \\ [0.0cm]
\epsfxsize=2.5in
\epsffile{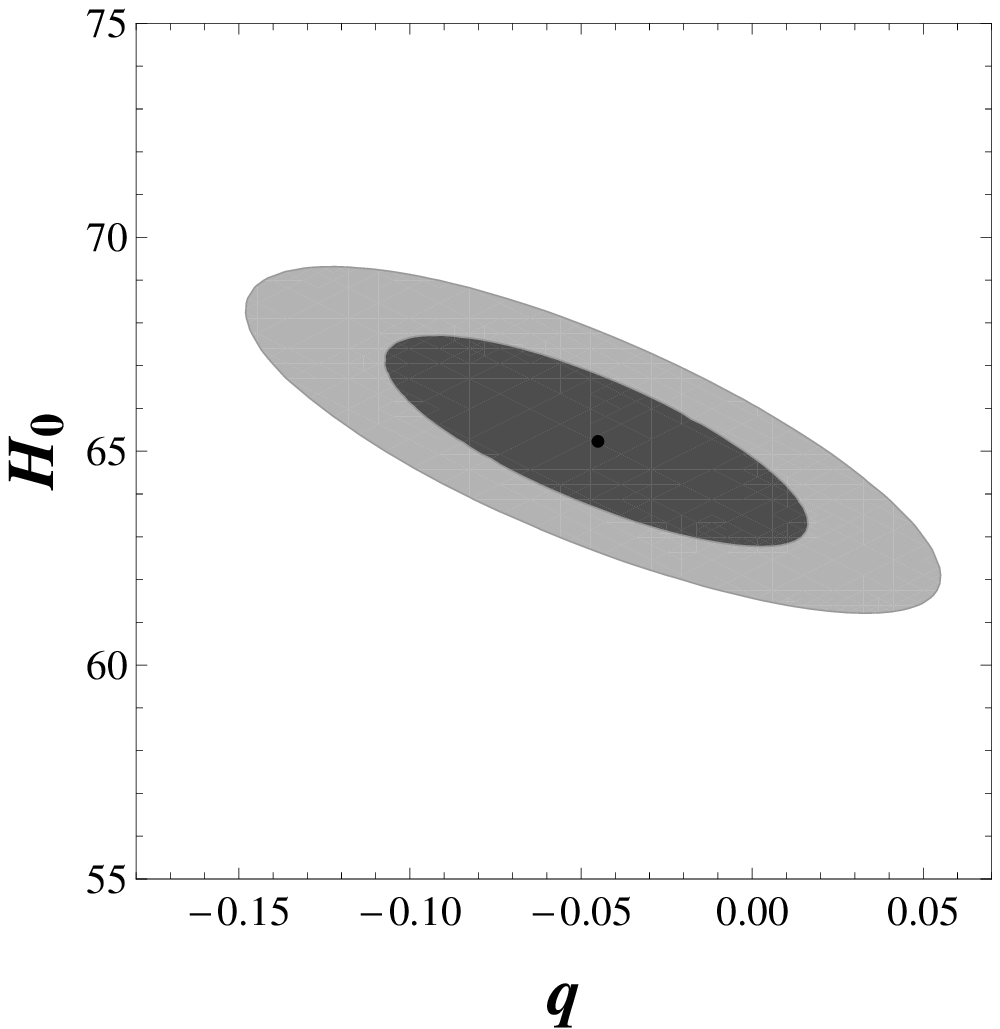} &
        \epsfxsize=2.6in
        \epsffile{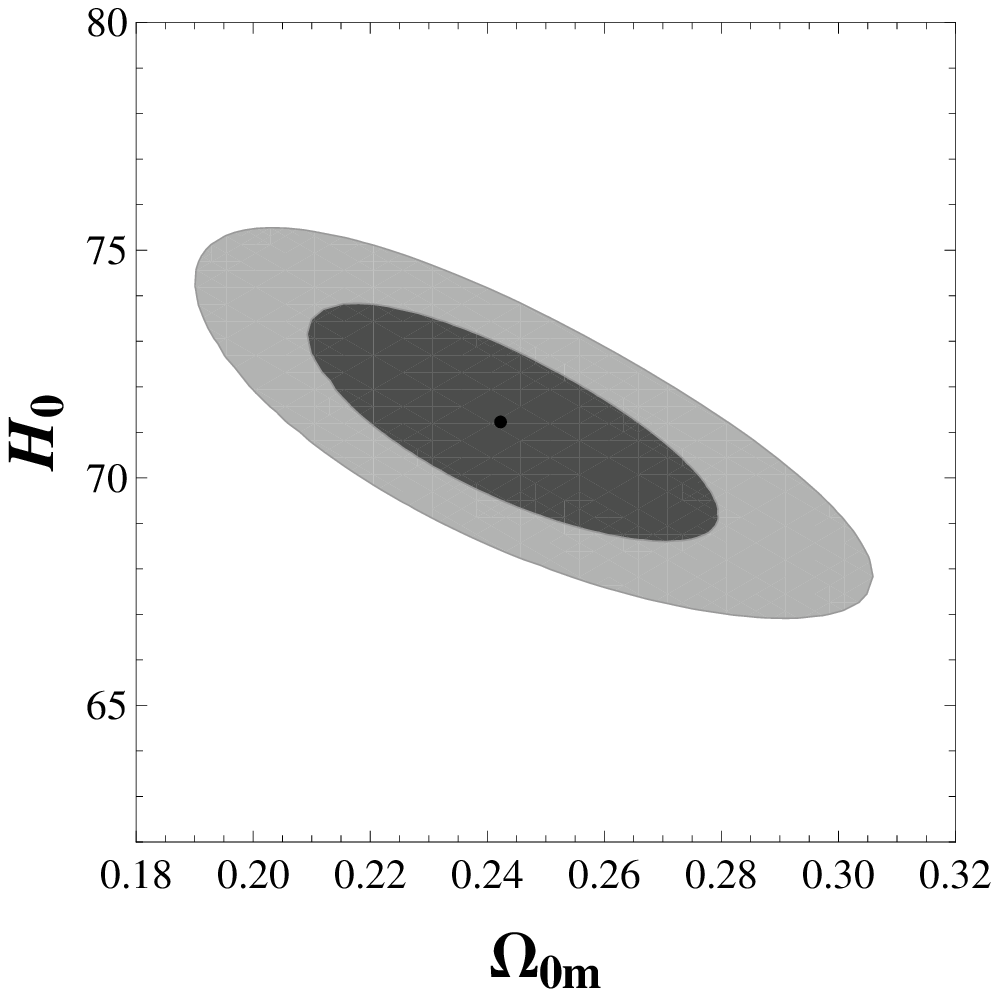} \\ [0.0cm]
\mbox{\bf (a)} & \mbox{\bf (b)}
\end{array}$
\end{center}
\caption{ \small This figure corresponds to the latest H(z) data. The panels (a) and (b) show the $1\sigma$ (dark shaded) and $2\sigma$ (light shaded) likelihood contours in the $q-H_{0}$ and $\Omega_{0m}-H_{0}$ planes, obtained for power-law and $\Lambda$CDM models respectively. The $H_0$ is represented in unit of Km/s/Mpc. Black dots designate the best fit values of the respective parameters.}
\label{hsn}
\end{figure*}
%%%%%%%
%%%%
\begin{figure*} \centering
\begin{center}
$\begin{array}{c@{\hspace{0.4in}}c}
\multicolumn{1}{l}{\mbox{}} &
        \multicolumn{1}{l}{\mbox{}} \\ [0.0cm]
\epsfxsize=2.55in
\epsffile{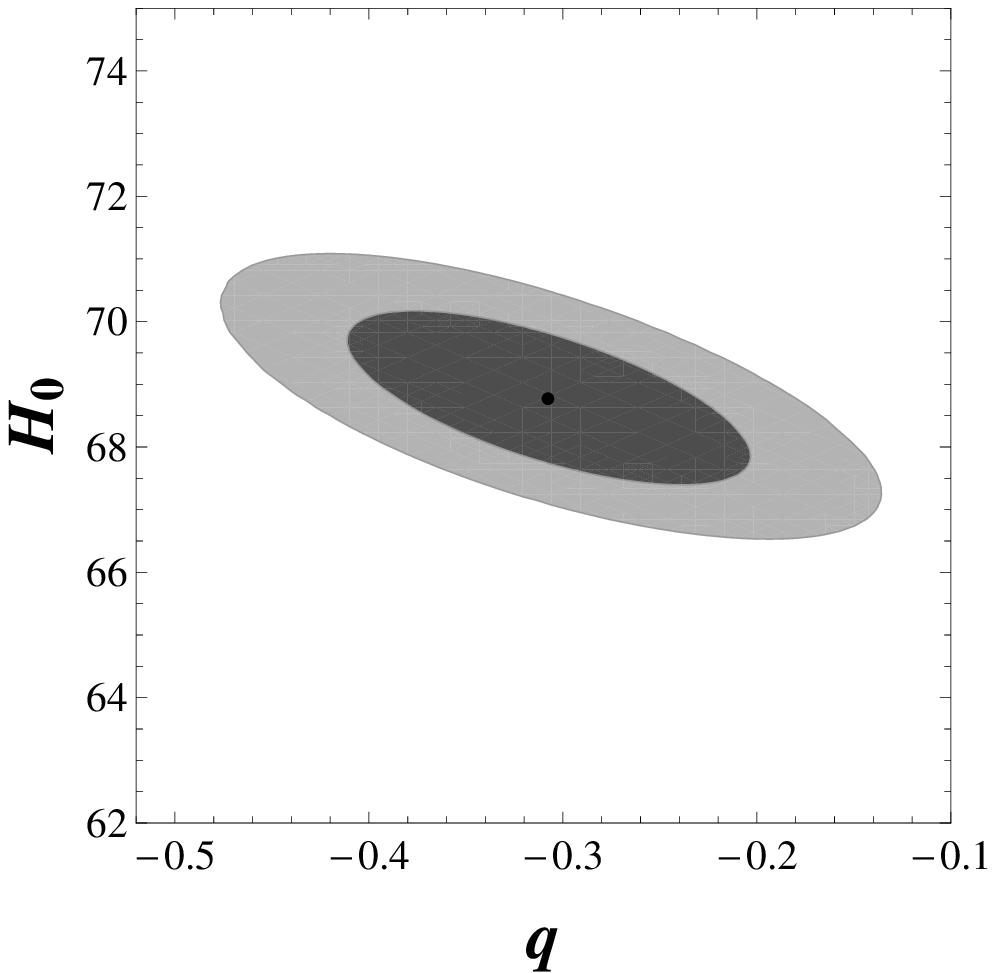} &
        \epsfxsize=2.44in
        \epsffile{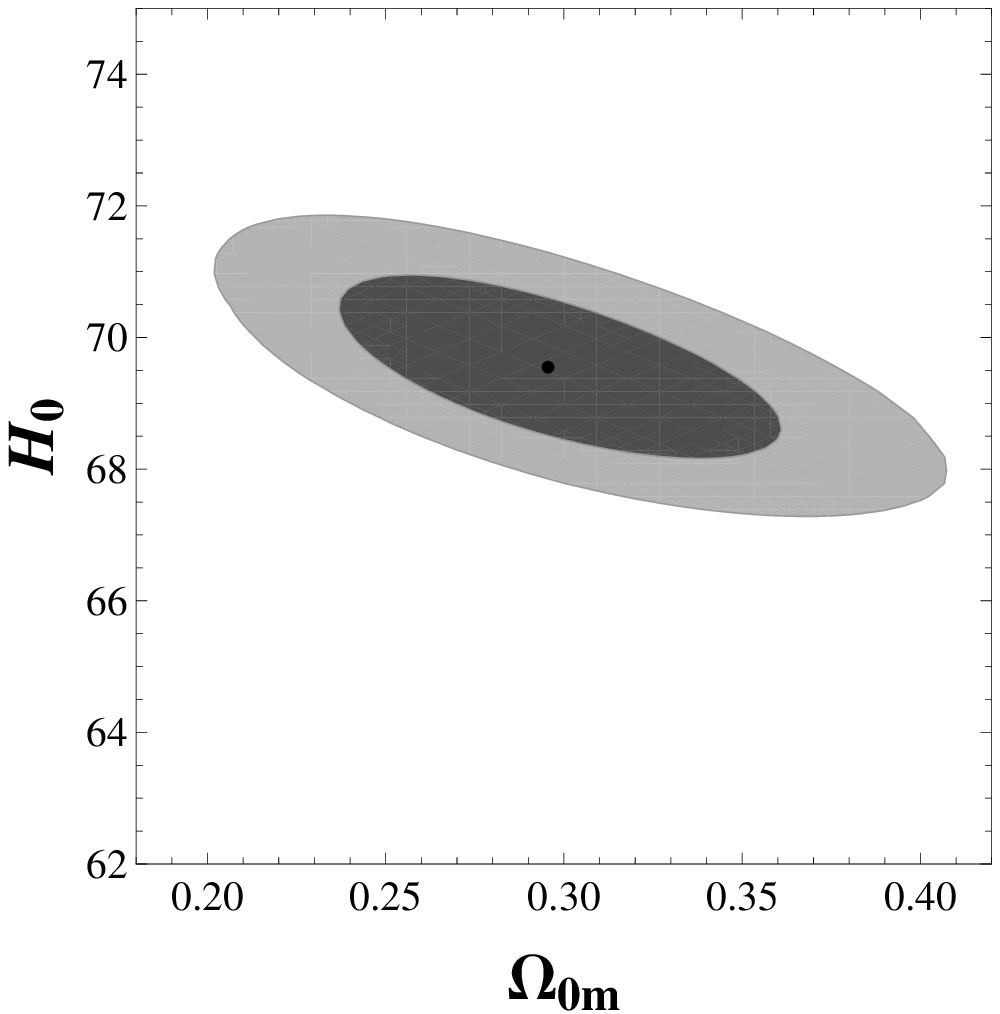} \\ [0.0cm]
\mbox{\bf (a)} & \mbox{\bf (b)}
\end{array}$
\end{center}
\caption{ \small This figure corresponds to SN data. The panels (a) and (b) show the $1\sigma$ (dark shaded) and $2\sigma$ (light shaded) likelihood contours in the $q-H_{0}$ and $\Omega_{0m}-H_{0}$ planes, obtained for power-law and $\Lambda$CDM models respectively. Here, the unit of $H_0$ is Km/s/Mpc. Black dots represent the best fit values of the respective parameters.}
\label{SNcont}
\end{figure*}
%%%%%%%
%%%%%%%%%%%
\begin{figure*} \centering
\begin{center}
$\begin{array}{c@{\hspace{0.4in}}c}
\multicolumn{1}{l}{\mbox{}} &
        \multicolumn{1}{l}{\mbox{}} \\ [0.0cm]
\epsfxsize=2.5in
\epsffile{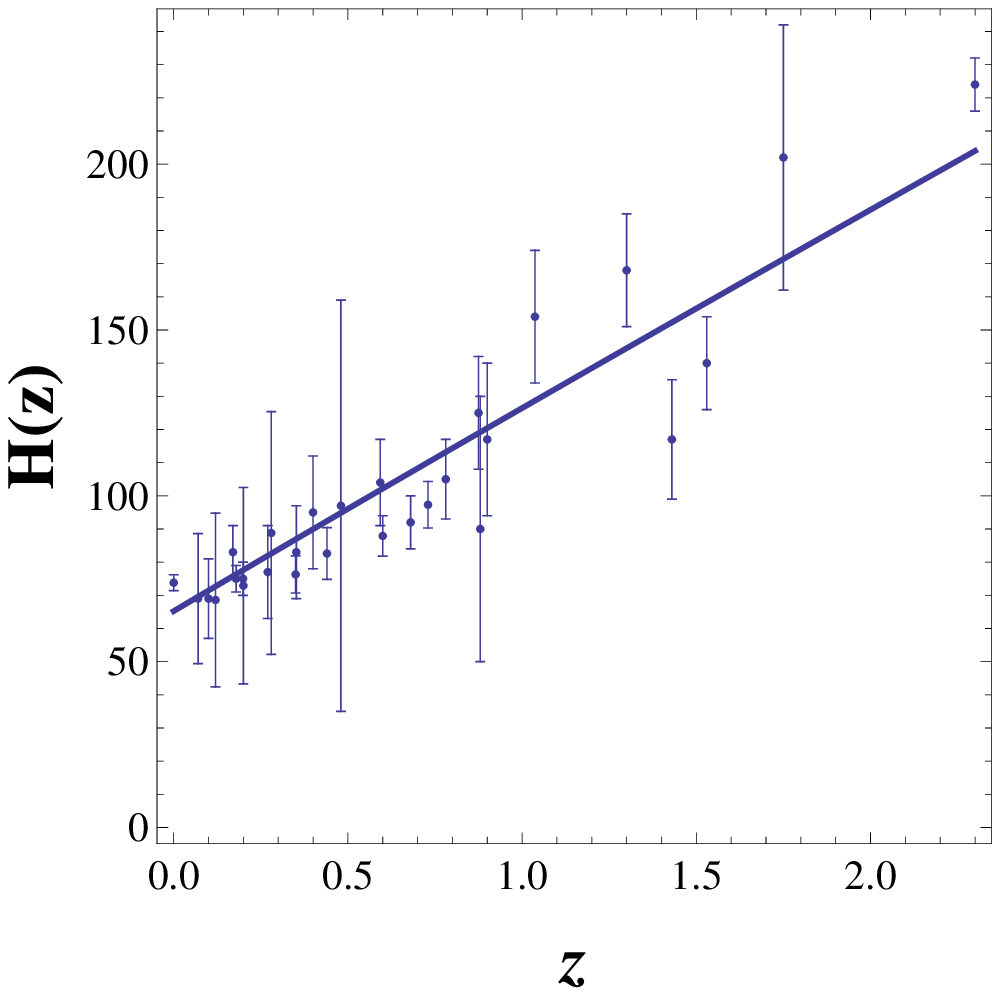} &
        \epsfxsize=2.44in
        \epsffile{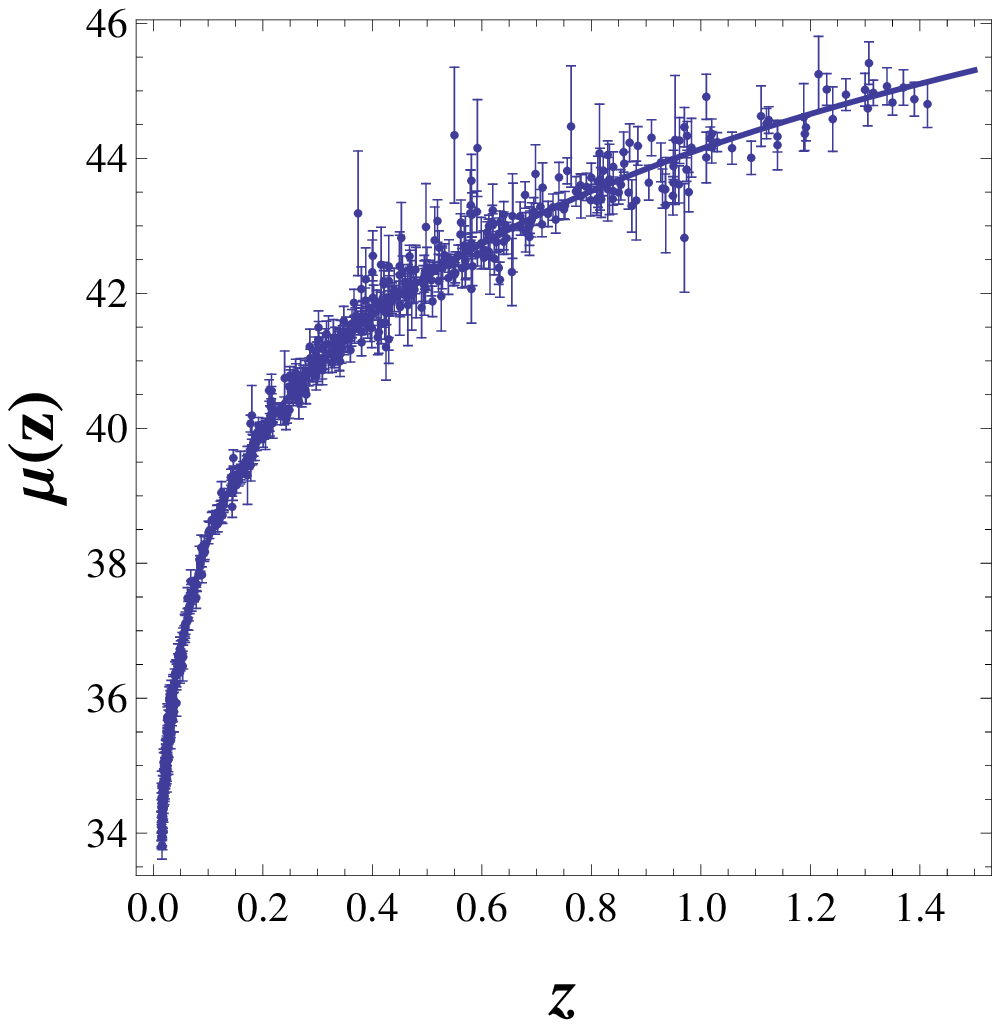} \\ [0.20cm]
\mbox{\bf (a)} & \mbox{\bf (b)}
\end{array}$
\end{center}
\caption{ \small Panels (a) and (b) correspond to the latest H(z) data and SN data respectively, with error bars. In both the panels, solid line shows the best fitted behaviour for power law cosmological model. The H(z) is expressed in unit of Km/s/Mpc. For panel (b), we have used diagonal covariance matrix.}
% The observational 15 $H(z)$ data points are shown with error bars (red color online). Variation of best fit model $H(z)$ curve (solid) based on $H(z)$+SNe Ia data is shown vs $z$. The dashed curve corresponds to the maximum values of $H(z)$ in the 1$\sigma$ region $68.41 \leq H_{0}\leq 69.46, -0.39\leq q \leq -0.29$ while the dotted curve corresponds to the minimum values of $H(z)$ in the same region. We observe that the best fit model fits well to the observational data points of $H(z)$ especially at redshits $z<1$

%%%%
%In panel (a)/(b), the observational 28 H(z)/580 SNe Ia data points are shown with error bars. Best fit model H(z)/$\mu(z)$ curve based on joint data i.e. (H(z)+ SNe Ia) versus redshift $z$ is shown by solid curve, whereas the dashed and dotted curve corresponds to the maximum and minimum values of H(z)/$\mu(z)$ in the $1\sigma$ region. In both of the panels, we see that  model fits well to  the  observational H(z)/SNe Ia data especially at redshift $z<1$.
\label{erbar}
\end{figure*}
%%%

\begin{table}
\caption{ $H(z)$ measurements (in unit [$\mathrm{km\,s^{-1}Mpc^{-1}}$]) and their errors \cite{Farooq:2013hq}.}
\begin{center}
\label{hubble}
\begin{tabular}{cccc}
\hline\hline
~~$z$ & ~~~~$H(z)$ &~~~~ $\sigma_{H}$ & ~~ Reference\\
~~    &~~ (km/s/Mpc) &~~~~ (km/s/Mpc)& \\
%\tableline
0.070&~~    69&~~~~~~~  19.6&~~ \cite{Zhang:2012mp}\\
0.100&~~    69&~~~~~~~  12&~~   \cite{Simon:2004tf}\\
0.120&~~    68.6&~~~~~~~    26.2&~~ \cite{Zhang:2012mp}\\
0.170&~~    83&~~~~~~~  8&~~    \cite{Simon:2004tf}\\
0.179&~~    75&~~~~~~~  4&~~    \cite{Moresco:2012by}\\
0.199&~~    75&~~~~~~~  5&~~    \cite{Moresco:2012by}\\
0.200&~~    72.9&~~~~~~~    29.6&~~ \cite{Zhang:2012mp}\\
0.270&~~    77&~~~~~~~  14&~~   \cite{Simon:2004tf}\\
0.280&~~    88.8&~~~~~~~    36.6&~~ \cite{Zhang:2012mp}\\
0.350&~~    76.3&~~~~~~~    5.6&~~  \cite{Chuang2012b}\\
0.352&~~    83&~~~~~~~  14&~~   \cite{Moresco:2012by}\\
0.400&~~    95&~~~~~~~  17&~~   \cite{Simon:2004tf}\\
0.440&~~    82.6&~~~~~~~    7.8&~~  \cite{Blake12}\\
0.480&~~    97&~~~~~~~  62&~~   \cite{Stern:2009ep}\\
0.593&~~    104&~~~~~~~ 13&~~   \cite{Moresco:2012by}\\
0.600&~~    87.9&~~~~~~~    6.1&~~  \cite{Blake12}\\
0.680&~~    92&~~~~~~~  8&~~    \cite{Moresco:2012by}\\
0.730&~~    97.3&~~~~~~~    7.0&~~  \cite{Blake12}\\
0.781&~~    105&~~~~~~~ 12&~~   \cite{Moresco:2012by}\\
0.875&~~    125&~~~~~~~ 17&~~   \cite{Moresco:2012by}\\
0.880&~~    90&~~~~~~~  40&~~   \cite{Stern:2009ep}\\
0.900&~~    117&~~~~~~~ 23&~~   \cite{Simon:2004tf}\\
1.037&~~    154&~~~~~~~ 20&~~   \cite{Moresco:2012by}\\
1.300&~~    168&~~~~~~~ 17&~~   \cite{Simon:2004tf}\\
1.430&~~    177&~~~~~~~ 18&~~   \cite{Simon:2004tf}\\
1.530&~~    140&~~~~~~~ 14&~~   \cite{Simon:2004tf}\\
1.750&~~    202&~~~~~~~ 40&~~   \cite{Simon:2004tf}\\
2.300&~~    224&~~~~~~~ 8&~~    \cite{Busca12}\\

\hline\hline
\end{tabular}
\end{center}
\end{table}
%%%%%%%%
\begin{table}
%\textbf{Table 3.} Summary of the parameters for flat universe.\\
\caption{ Summary of the numerical results  for flat power law cosmological model.}
%\begin{ruledtabular}
\begin{center}
\label{tabparm}
\begin{tabular}{l c c c r}
\hline\hline
Data  &$q$  &$H_0${\footnotesize(km/s/Mpc)} & $\beta$ &Refs.
\\
\hline
\\
$H(z)$ {\footnotesize{(14 points)} }   &  $-0.18^{+0.12}_{-0.12}$  &   $68.43^{+2.84}_{-2.80} $     &  $ -$   &
\\
\\
SN {\footnotesize{(Union2)} }   &  $-0.38^{+0.05}_{-0.05}$  &   $69.18^{+0.55}_{-0.54} $     &  $ -$   & \footnotesize{\cite{sur}}
\\
\hline
\\
WMAP7   &  $ -$    &   $70.3^{+2.5}_{-2.5}$     &   $ 0.99^{+0.04}_{-0.04}  $    &\footnotesize{\cite{gum}}
\\
\\
WMAP7+BAO+$H(z)$ &  $ -$  & $70.4^{+1.4}_{-1.4}$  &  $  0.99^{+0.02}_{-0.02}   $
\\
\hline
\\
$H(z)$ {\footnotesize{(29 points)}}  &  $ -0.0451_{-0.0625}^{+0.0614}$             &   $65.2299_{-2.4607}^{+2.4862}$     &   $ -$
\\
\\
SN {\footnotesize{(Union2.1 )} } &   $-0.3077_{-0.1036}^{+0.1045}$  & $68.7702_{-1.3754}^{+1.4052}$  &  $ -$  & \footnotesize{This Letter}
\\
\\
\hline\hline
\end{tabular}
%\end{ruledtabular}
\end{center}
\end{table}
%%%%%%%%%
\begin{table}
%\textbf{Table 2.} Summary of the numerical results.\\
\caption{ Numerical results summary of simulated data.}
\begin{center}
\label{tabjdem}
\begin{tabular}{l c c c c r }
\hline\hline
Data & $<q>$ & $<H_{0}>${\footnotesize(km/s/Mpc)} & $<r>$& $<s>$\\
\hline
\\
JDEM & $-0.0656_{+0.0003}^{-0.0003}$ & $25.9083_{+0.0030}^{-0.0030}$ & $-0.0570_{+0.0002}^{-0.0002}$ & $0.6229_{+0.0002}^{-0.0002}$ &  \\\\
\hline\hline
\end{tabular}
\end{center}
\end{table}
%%%%%%%%%%%
\section{$\Lambda$CDM}
As per the point of view of cosmology, the most simplest candidate of dark energy is the cosmological constant whose energy density remains constant with time i.e. $\rho_\Lambda \equiv \frac{\Lambda}{8\pi G} = -p_\Lambda $ and its equation of state is, $\omega_\Lambda=-1$. A Universe having matter in the form of dust and $\Lambda$ is known as $\Lambda$CDM. In flat FLRW Universe, Hubble parameter for $\Lambda$CDM model has the following form:

\beq\label{c1}
H(z)=H_0~ [\Omega_{0m} (1+z)^3+(1-\Omega_{0m})]^{1/2}.
\eeq
Here, $\Omega_{0m}$ and $H_0$ are the present matter density and Hubble parameters respectively. 

\section{Observational Constraints}

\begin{itemize}
  \item \textbf{H(z) Data:}\\
  We find the observational constraints on both of the parameters $H_0$ and $q$ to the latest $28$ data points of H(z) \cite{Farooq:2013hq} in the redshift range $0.07\leq z \leq 2.3 $. The values are presented in the Table \ref{hubble}. To complete the data set, we use $H_0$, as estimated in the reference \cite{rie}.\\
we define $\chi^2$ as
  \beq\label{h1}
  \chi^2_{H}=\sum_{j=1}^{29}\frac{(H_{exp}(z_{j})-H_{obs}(z_{j}))^2}{\sigma_{j}^2}.
  \eeq

Where $H_{exp}$ is the expected value of the Hubble parameter, $H_{obs}$ is the observational value and, $\sigma_j$ is the corresponding $1 \sigma$ error. The power law cosmological model contains  two independent parameters namely $q$ and $H_{0}$. As $\beta>0$ is required in power law cosmology, hence $q>-1$ and $H_0\geq 0$, therefore we find the best fit values of $q$ and $H_0$ by restricting the parametric space  as $q>-1$ and $H_{0}\geq0$. As a result, we obtain the best fit values of the parameters as $q=-0.0451$, $H_{0}=65.2299$ km/s/Mpc and  $\chi^2_{\delta}=1.8131$, and the values of the parameters with $1\sigma$ error are obtained as  $q=-0.0451_{-0.0625}^{+0.0614}$ and $H_{0}=65.2299_{-2.4607}^{+2.4862}$ km/s/Mpc, where $\chi^2_{\delta}=\chi^2_{min}$/(degree of freedom). We carry out the same analysis for $\Lambda$CDM model and as a result, the best fit values of the parameters are obtained as $\Omega_{0m}=0.2422$, $H_0=71.2272$ km/s/Mpc and  $\chi^2_{\delta}=0.7174$, and the values of the parameters with $1\sigma$ error are obtained as  $\Omega_{0m}=0.2422_{-0.0331}^{+0.0373}$ and $H_{0}=71.2272_{-2.6291}^{+2.6112}$ km/s/Mpc. By comparing the values of $\chi^2_{\delta}$ of both the models, we find that the power law cosmological model does not fit well to the latest H(z) data. The $1\sigma$ (dark shaded) and $2\sigma$ (light shaded) likelihood contours for both the models are shown in  figure \ref{hsn}. Best fitted behaviour for power law cosmological model with H(z) data and error bars has been shown in panel (a) of figure \ref{erbar}. It is worthwhile to notice from Table \ref{tabparm} that the best fit value of $q=-0.18$ \cite{sur} has been significantly shifted to $q=-0.0451$ with latest H(z) data. It was claimed in reference \cite{sur} that power law model fits well with H(z) data but in our case, by comparing the values of $\chi^2_{\delta}$ of power law and $\Lambda$CDM models, we find that the power law cosmological model does not fit well with the latest H(z) data.
{\bf }
%%%%%%%%
\item \textbf{ Union2.1 SN Data :}\\
We now put constraints on the above said parameters by using Type Ia supernova observation which is one of the direct probes for the cosmological expansion. SNe Ia are always used as standard candles for estimating the apparent magnitude $m(z)$ at peak brightness after accounting for various corrections, and are believed to provide strongest constraints on the cosmological parameters. In this investigation, we work with recently released Union2.1 compilation  set of $580$ SNe Ia data points. For a standard candle of absolute magnitude $M$ and luminosity distance $d_{L}$, the apparent magnitude $m(z)$ is expressed as
\beq\label{s1}
m=M+5\log_{10}\left(\frac{d_{L}}{1\;Mpc}\right)+25,
\eeq
where $M$ is constant for all SNe Ia. Equation (\ref{s1}) can be written as
\beq\label{s2}
m=M+5\log_{10}D_{L}(z)-5\log_{10}H_{0}+52.38\;\;.
\eeq
where
\beq\label{s3}
D_{L}(z)=\frac{H_{0}}{c}d_{L}(z).
\eeq

The distance modulus $\mu(z)=m-M$ is given by
\beq\label{s4}
\mu(z)=5\log_{10}D_{L}(z)-5\log_{10}H_{0}+52.38\;\;.
\eeq
$D_{L}$, the Hubble free luminosity distance can now be expressed as
\beq\label{s5}
D_{L}(z)=(1+z)\int_{0}^{z}\frac{H_{0}}{H(z^*)}dz^*=\frac{1}{q}\left[(1+z)-\frac{1}{(1+z)^{q-1}}\right].
\eeq
 $\chi^2$ can be defined as
\beq\label{s6}
\chi^2_{SN}=\sum_{ij}{\Big (}\mu(exp)_i-\mu(obs)_i{\Big )}C_{ij}^{-1}{\Big (}\mu(exp)_j-\mu(obs)_j{\Big )},
\eeq
where $C_{ij}$ is the full covariance matrix \cite{Suzuki:2011hu}.
Best fit values of the power law model parameters in the parametric space ($q>-1$ and $H_{0}\geq0$) with $1\sigma$ error are obtained as  $q=-0.3077_{-0.1036}^{+0.1045}$ and $H_{0}=68.7702_{-1.3754}^{+1.4052}$ km/s/Mpc together with $\chi^2_{\delta}=0.9464$. Similar results obtained for $\Lambda$CDM model parameters are $\Omega_{0m}=0.2955_{-0.0587}^{+0.0655}$ and $H_{0}=69.5493_{-1.3885}^{+1.4039}$ km/s/Mpc together with $\chi^2_{\delta}=0.9431$. Here, we observe that values of $\chi^2_{\delta}$ of both the models are approximately equal. Therefore, we can say that power law cosmological model fits well with Union$2.1$ compilation data. The $1\sigma$ (dark shaded) and $2\sigma$ (light shaded) likelihood contours for both the models are shown in  figure \ref{SNcont}. Best fitted behaviour for power law cosmological model with SN data and error bars has been shown in panel (b) of figure \ref{erbar}. We also observe from Table \ref{tabparm} that the revised analysis of SN data gives much larger error bars on both $q$ and $H_0$ than that of reference \cite{sur}. This might have occurred due to taking into account the full covariance matrix of the latest SN data in the present analysis that was ignored in reference \cite{sur}.

~~~From the best fit values presented for power law model with H(z) data and SN data, we observe that there is a large discrepancy between the values of $q$, this might occur because in the present updated observational analysis we see that power law model fits well with SN data but not with H(z) data.
%%%
\item \textbf{JDEM: Simulated SN data }\\
We use simulated dataset \cite{hol} for the future JDEM, supernova surveys having approximately 2300 supernovae in the redshift range 0 $-$ 1.7. The errors do not depend on redshift and equal for all SNe. We take $\sigma = 0.13$ \cite{ald}.
%To look at what can be achieved in future, we also use a simulated dataset \cite{hol} based on the upcoming  JDEM SN-survey containing around $2300$ SNe. These are distributed over a redshift range from $z=0$ to $z=1.7$. We assume a simplified error model where the errors are  same for all supernovae and independent of redshift. We assume a statistical error of $\sigma = 0.13$ magnitude, as expected from JDEM-like future surveys \cite{ald}. 
Using latest specifications, we generated $500$ data sets as explained in \cite{alam2003}, considering $\Lambda$CDM model as our fiducial model with $\Omega_{0m} = 0.3$. For each of these experiments i.e. $500$ experiments, the best-fitting parameters $H_0$ and $q$ were calculated. We then calculated $r$ and $s$ for each experiment from the calculated values of the model parameters and finally we computed the mean values of $H_0$, $q$, $r$, and $s$ as $<H_0>$, $<q>$, $<r>$, and $<s>$. The numerical results  obtained have been presented in Table \ref{tabjdem}.
\end{itemize}
%%%%%%%%%
\subsection{Constraints on Statefinders}
\begin{figure*} \centering
\begin{center}
$\begin{array}{c@{\hspace{0.4in}}c}
\multicolumn{1}{l}{\mbox{}} &
        \multicolumn{1}{l}{\mbox{}} \\ [0.0cm]
\epsfxsize=2.5in
\epsffile{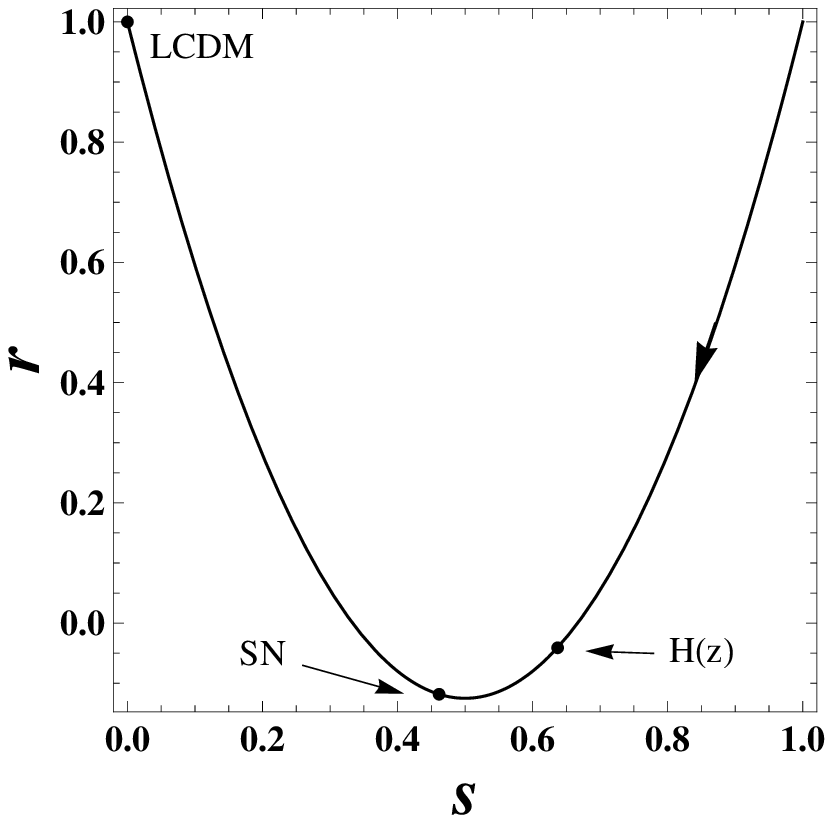} &
        \epsfxsize=2.44in
        \epsffile{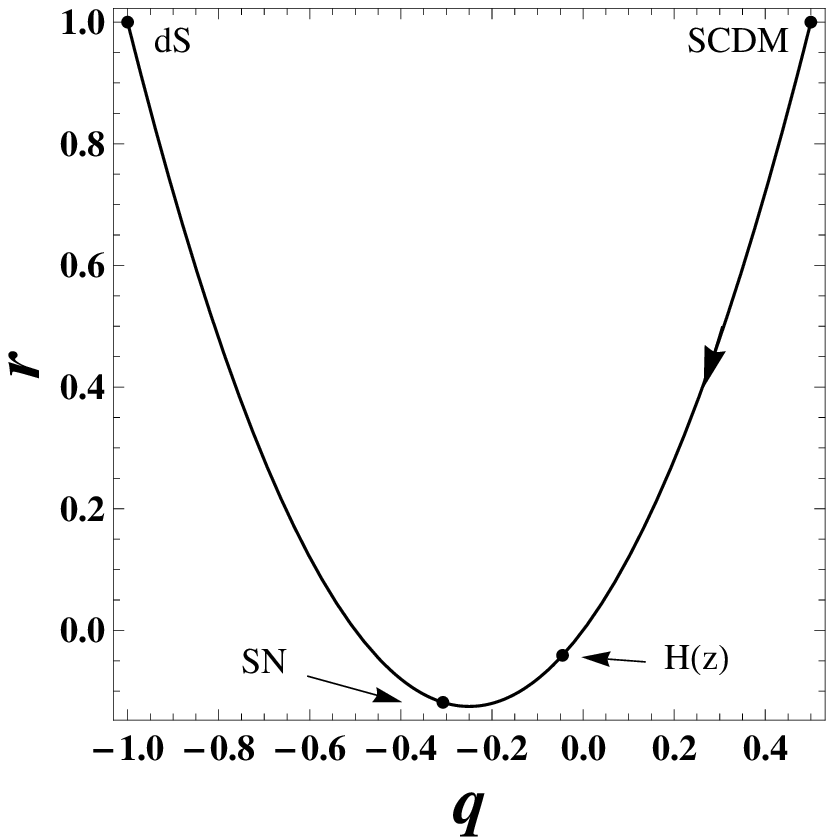} \\ [0.20cm]
\mbox{\bf (a)} & \mbox{\bf (b)}
\end{array}$
\end{center}
\caption{ \small  The panel (a)  shows the time evolution of the statefinder pair $\lbrace r,s \rbrace$ for power- law cosmological model. The model converge to the fixed point ($r=1, s=0$) which corresponds to LCDM. The panel (b) shows the time evolution of the statefinder pair $\lbrace r,q\rbrace$ for said  model. The  point ($r=1,q=0.5$) corresponds to a matter dominated Universe (SCDM) and converges to the point ($r=1,q=-1$) which corresponds to the de Sitter expansion (dS). The black dots shown on the curves by arrows are the best fit values of $r$, $s$ and $q$ obtained by latest H(z) and SN data.}
\label{rsq}
\end{figure*}
%%%
\begin{table}
%\textbf{Table 4.} Constraints on statefinders from H(z), SN and joint data\\
\caption{Constraints on statefinders from latest H(z) and  SN data.}
\begin{center}
\label{tabstat}
\begin{tabular}{l c c c r}
\hline\hline
Data & $r$& $s$ & Refs.
\\
\hline
\\
$H(z)$ {\footnotesize{(14 points)} }   &  $-0.09^{+0.04}_{-0.03}$  &   $0.58^{+0.04}_{-0.12} $     &
\\
\\
SN {\footnotesize{(Union2)} }   &  $-0.09^{+0.03}_{-0.02}$  &   $0.41^{+0.03}_{-0.03} $     &   \footnotesize{\cite{sur}}
\\
\hline
\\
$H(z)$ {\footnotesize{(29 points)} } & $-0.0410_{-0.0512}^{+0.0504}$ & $0.6366_{-0.0417}^{+0.0409}$
\\\\
SN {\footnotesize{(Union2.1 )} } &  $-0.1183_{-0.0241}^{+0.0239}$ & $0.4615_{-0.0691}^{+0.0697}$  & \footnotesize{This Letter}
\\\\
\hline\hline
\end{tabular}
\end{center}
\end{table}
%%%%
Following the analysis presented in reference [13], we obtain constraints on the statefinders with latest H(z) data as $r=-0.0410_{-0.0512}^{+0.0504}$, $s=0.6366_{-0.0417}^{+0.0409}$ and, with SN data as $r=-0.1183_{-0.0241}^{+0.0239}$ and $s=0.4615_{-0.0691}^{+0.0697}$. The above results have been summarized in  Table \ref{tabstat}.

We observe that power law model does not fit well with latest H(z) data and therefore significant changes in the best fit values are noticed from Table \ref{tabparm} and the corresponding variations in the constraints on statefinders can be observed from Table \ref{tabstat} but it fits well with updated SN data as well as previous data \cite{sur} therefore marginal differences are observed in the constraints from Table \ref{tabparm} and similar kind of variations can be noticed in statefinders from Table \ref{tabstat}.

\section{Conclusion}
Precision cosmological observations offer
the possibility of uncovering essential
properties of the Universe. Here, we have
investigated power-law cosmology $a(t) \, \propto t^{\beta}$,
which has some prominent features, making it unique when compared
to other models of the Universe. For example, for $\beta\geq1$,
it addresses to the horizon, flatness and age problems \cite{kolb, mann, allen}
and all these features provide viability to the power-law cosmology to dynamically
solve the cosmological constant problem. In the work presented here, we used the most
 recent observational data sets from H(z) and SNe Ia observations and obtained the constraints on the two crucial cosmological parameters $H_0$ and $q$ and compared our results with reference \cite{sur}.
 We also have forecasted these constraints with simulated data for large future
  surveys like JDEM. Statistically, this model may be preferred over other models
  as we have to fit only two parameters. Numerical results obtained have been  concluded in the Tables \ref{tabparm}, \ref{tabjdem} and \ref{tabstat}.
  
In this work, we observed that though $q$ is negative in the constraints from both H(z) and SNe Ia observations respectively but with bad $\chi_{\delta}^2$ in case of H(z) data, thus we can say that latter explains the present cosmic acceleration more efficiently than that of former in the context of power law cosmology. With the latest H(z) data, we found that obtained best fit value of $H_0$ for power law is outside more than 2$\sigma$ confidence level from the value of $\Lambda$CDM and also, we noticed the large discrepancy between the values of their $\chi^2_{\delta}$.  In contrast, with SN data, we found that the value of $H_0$ for power law agrees with $\Lambda$CDM within 1$\sigma$ confidence level and also, its $\chi^2_{\delta}$ is approximately equal to the value of $\chi^2_{\delta}$ for $\Lambda$CDM. Therefore, we conclude that power law model fits well  with SN data but not with H(z) data. Contour plots for both the models with H(z) data and SN data have been shown in figures \ref{hsn} and \ref{SNcont} respectively. Best fitted behaviour for power law model with data error bars have been shown in figure \ref{erbar}. On comparing our results with reference \cite{sur}, we observe that in the new analysis, best fit value of $q$ with H(z) data is drastically different from the constraint with SN data but it was not so significant in reference \cite{sur}. This discrepancy has been observed because in the current analysis, power law model does not fit well with latest H(z) data due to significant difference between the value of its $\chi^2_{\delta}$ and $\Lambda$CDM. Also we observe that new SN analysis gives larger error bars on both $q$ and $H_0$ than that of reference \cite{sur} because of taking into account the full covariance matrix. Corresponding variations in the values of statefinders have also been observed which have been summarized in Table \ref{tabstat}.

More explicitly, we see the differences in our study and of reference \cite{sur} as: in latter one it had been shown that the power law model fits well with both H(z) and SNe Ia observations but in our case we observed that it fits well only with SN data having larger error bars on both of the parameters and on the contrary it fails to fit with latest H(z) data shifting best fit value of $q$ significantly with bad $\chi_{\delta}^2$. Thus, we can say that our study explains merits and demerits of power law model in explaining the evolution of Universe in a more clear and sophisticated manner than that of reference \cite{sur}. The statefinder diagnostic carried out shows that power law cosmological model will finally approach  the $\Lambda$CDM model as shown in figure \ref{rsq}.
   From the results mentioned in Table \ref{tabjdem}, one can also conclude that future surveys like JDEM
   demands an accelerated expansion of the Universe but with smaller values of Hubble constant
   within the framework of power law cosmology. From the above discussed results, it can be concluded that though power law cosmology has several prominent features but still it fails to explain redshift based transition of the Universe from deceleration to acceleration, because here we do not have redshift or time dependent deceleration parameter $q$. Thus, in nutshell it can clearly be said that despite having numerous remarkable features, the power law cosmology does not fit well in dealing with all cosmological challenges.
%%%
%\begin{table*}
%\textbf{Table 3.} Summary of the numerical results\\

%\begin{tabular}{l c c c c c}
%\hline
%Data/Source & $q$ & $H_{0}$  & $\chi^2_{\nu}$& $r$& $s$\\
%\hline
%$H(z)$ & $-0.0440_{-0.0508}^{+0.0496}$ & $65.1738_{-1.5990}^{+1.6035}$ & $1.3509$ & $-0.0401_{-0.0419}%^{+0.0409}$ & $0.6373_{-0.0339}^{+0.0331}$ \\\\
 % SNe Ia & $-0.3610_{-0.0507}^{+0.0510}$ &$69.1659_{-0.5185}^{+0.5220}$& $0.9798$& $-0.1003_{-0.0226}%^{+0.0225}$ & $0.4260_{-0.0338}^{+0.0340}$   \\\\
% $H(z)+$SNe Ia& $q=-0.2086_{-0.0379}^{+0.0374}$ &  $68.0209_{-0.4436}^{+0.4477}$& $1.0683$& %%%$-0.1216_{-0.0063}^{+0.0062}$ & $0.5276_{-0.0253}^{+0.0249}$   \\\\
%\hline
%\end{tabular}
%\label{table:nonlin}
%\end{table*}
\section*{Acknowledgment}
We are indebted to M. Sami for useful discussions and comments. Author SR thanks  A. A. Sen, S. Jhingan and the whole CTP, JMI for providing the necessary facilities throughout this work. SR also acknowledges Gurmeet Singh and Vikas Kumar for their continuous support in improving this manuscript.

\end{document}